# Turing-type 3D tubular pattern formed by water molecules: A simulation of the ice crystal growth inside a sandwich structure from the perspective of reaction-diffusion mechanism

Xiaolu Zhu[1,2,3*] and Zheng Wang[1]

[1]College of Mechanical & Electrical Engineering, Hohai University, Changzhou Campus, Jiangsu 213022, China

[2]Changzhou Key Laboratory of Digital Manufacture Technology, Hohai University, Changzhou, Jiangsu 213022, China;

[3]Jiangsu Key Laboratory of Special Robot Technology, Hohai University, Changzhou Campus, Jiangsu 213022, China;

* Corresponding Author

Email: zhuxiaolu@hhu.edu.cn; Tel: +86-1586-186-3691

**Abstract:** The process of formation for an ice crystal with elaborate, symmetrical patterns from the water vapor is usually complicated and the corresponding mechanism is still not clear. Here, we experimentally constructed the 3D tubular ice crystals within a thin chamber layer filled with air in a freezer with high humidity at -27 ℃. We here also propose to investigate the dynamic formation of hollow structures from the perspective of Turing's reaction-diffusion process, and a mathematical modelling of the 3D tubular structures composed of ice crystals is conducted by regarding the cooled air molecule (C-AM) as the activator and room-temperature air molecules(RT-AM) as the inhibitor. The simulation generated a hollow tube array that has similar geometric feature with the ice crystal structure of hollow columns in the experiment. This model offers a possibility to explore the Turing instability in the process of ice crystal formation.

## I. INTRODUCTION

Ice formation is a common phenomenon and has widespread influence on fundamental research, industrial processes and daily life [1,2]. The snowflakes, also called snow crystals are single crystals of ice that grow from water vapor. The process of formation for an ice crystal (a snowflake) with elaborate, symmetrical patterns from the water vapor is usually complicated, yet it closely related two parameters: temperature (T) and water vapor supersaturation (σ) [3,4]. Nakaya first created artificial snow in the laboratory, and observed different growth patterns of snow crystals those determined by different temperatures and supersaturation levels, originating from which the famous "snow crystal morphology diagram" had been plotted [5,6]. The basic shape of a growing crystal depends on mainly on temperature while the appearance of ice structure depends more on the level of supersaturation. When the humidity is increasing, rapidly growing columns could become feathery needle crystals, and hexagonal plates grow into sectored plates, stellar dendrites or hollow columns [6]. The intrinsic growth rates of the crystal facets, and diffusion serves as the two main factors for the produced morphology of snow crystals.

Although the studies on ice crystal growth have obtained the many progresses, the driven force and the fundamental dynamics underlying the ice crystal growth are still not clear. Currently, one of the interesting points is how do the hollow columns forms, because most morphologies of ice crystals are not hollow. Recently, Zhang et al. [3] investigated the formation of hexagonal ice crystals from water vapor in a supersaturated water vapor environment by in situ observation via environmental scanning electron microscopy (ESEM), which confirms that supersaturated water vapor environment is required and the hexagonal ice crystals are developed by a step-by-step pathway including the screw dislocations and initial steps. However, it does not involve the description or speculation on the reason for the formation of hollow columns of ice crystals.

Previously, we investigated the tubular structure formation by cellular self-assembly via micro-engineering approaches [7,8]. The mechanism for the hollow tube formation via cellular self-assembly had been speculated by utilizing the theoretical model based on Turing's reaction-diffusion (RD) process [9]. Since the similar morphology of the self-assembled hollow structures composed of ice crystals or cells, we here propose to investigate the dynamic formation of hollow structures from the aspect of Turing's reaction-diffusion (RD) process whose foundation was described in the literatures [10-13]. In this study, the mathematical modelling of the 3D tubular structures composed of ice crystals is conducted by regarding the cooled air molecule (CM) as the activator and ambient air molecules(AM) as the inhibitor. The formation of ice crystal structure is hypothesized to depend on the reaction and diffusion process of the above activator and inhibitor.

## II. EXPERIMENTAL SETUP AND RESULTS

Here, we create the hollow ice tube composed of ice crystals in a freezer (BC-BD-102HT, Haier, China). The temperature (T) of the freezer was set at -27 ℃ and the humidity is around 95-99%. Water vapor pressure in the environment is becoming lower and lower when the T decreased. And it will be lower than the saturate water vapor pressure on the ice surface. Therefore, the inside of the freezer has a supersaturation condition of water vapor.

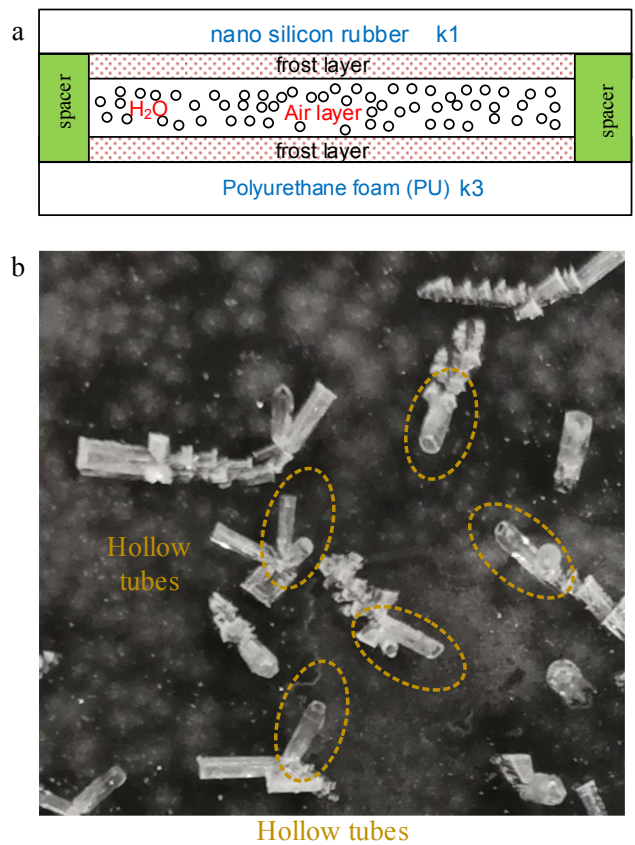

Figure 1. Experimental setup (a) and result (b) for creating the ice crystals inside a thin chamber layer in a freezer.

The bottom of the freezer is a polyurethane foam, based on which we built a thin chamber layer called air layer by using a nano silicon rubber to cover the polyurethane foam with tape spacers with around 4-5mm. The coefficients of thermal conductivity for different regions were set as $k_w$ = 0.55- 0.59 w/m·K for water vapor in the air layer and frost layer, and the $k_w$ may a little lower in the frost layer. $k_1$=0.15 - 0.27 w/m·K for the nano silicon rubber and $k_2$ = 0.021 w/m·K for the polyurethane foam. After 4-8 days, the ice crystals with hollow tube morphology emerged as shown in Fig. 2. The hollow ice tubes have an outer diameter of ~0.5-2.0 mm and a length of ~ 2-5mm and they attached on the rubber plate surface.

## III. THEORETICAL MODEL AND SIMULATIONS

Self-organization of water molecules also exists in the atmosphere in the closed chamber of the freezer. Here, the density of the water molecule in the 3D air layer and frost region in the serve as a main study object. The reaction-diffusion of morphogens consisting activator and inhibitor are modelled. The thermal energy distributed unevenly within this 3D space in this refrigerating system. It is hypothesized that the thermal energy at an arbitrary position (x,y,z) could be determined/described by density of cooled air molecule (C-AM) and density of room-temperature air molecules (RT-AM) which could be noted as U (x,y,z) and V(x,y,z). The C-AMs indicated the air molecules already cooled to the set temperature by the freezer, and they have lower kinetic energy because the heat have been partially removed from them; while the ambient air molecules or called room-temperature air molecules (RT-AM) have higher kinetic energy because they have not been cooled yet. The C-AM and RT-AM have the opposite thermal effect, and inhibit each other. The density of heat energy at a certain position (x,y,z) should be calculated by $k_1V(x,y,z) - k_2U(x,y,z)$. The combinational transfer of C-AM and RT-AM within 3D space conjunctly determines the distribution of the thermal energy. In this Turing-type model, the C-AM serves as the slowly-diffusing activators while the RT-AM serves as its rapidly-diffusing inhibitor. The C-AMs have lower temperature and lower kinetic energy and thus diffuse slow. The RT-AMs have higher temperature and higher kinetic energy, and thus diffuse rapidly. The freezer's cooling process usually slower and more arduous than the temperature recovery process from low temperature to ambient temperature, which is also coincident with above statements.

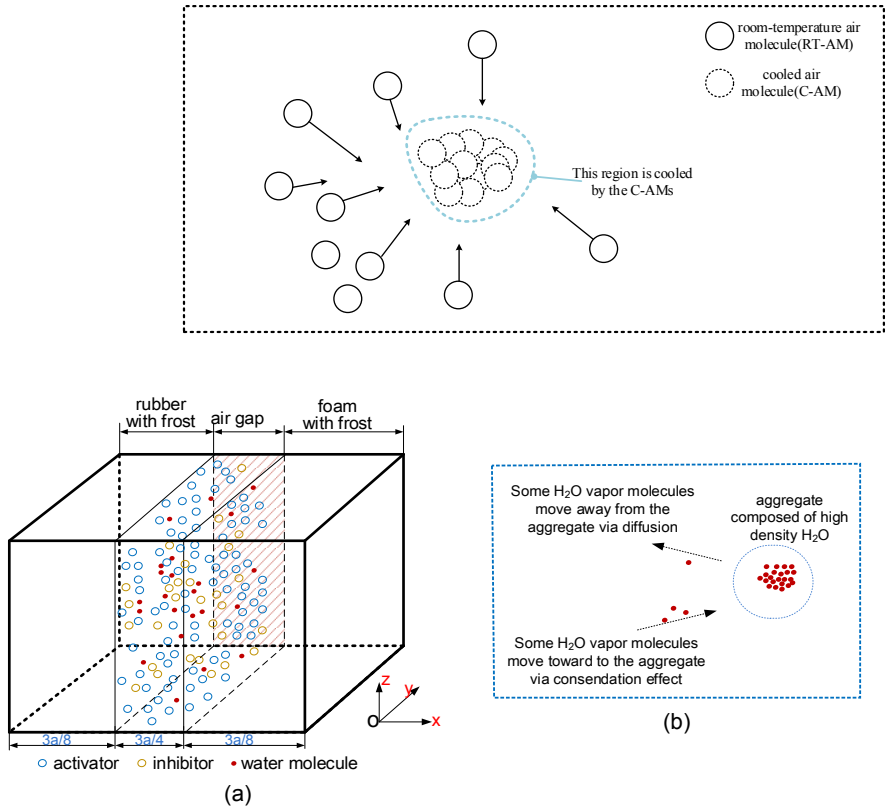

Figure 2. The schematic of the proposed reaction-diffusion system of cooled air molecule (C-AM), ambient air molecules (RT-AM) and gaseous water molecules.

Based on our previous study [12], We modeled the air-H$_2$O cooling system as the reaction (using Gierer and Meinhardt kinetics) and diffusion of the autocatalytic and slowly-diffusing equivalent activator, U, its rapidly-diffusing equivalent inhibitor, V, and H$_2$O density, n, reflecting air molecule diffusion, thermo-tactic migration and ice crystal growth with respect to activators U, as functions of a 3-dimensional domain (x, y, z):

$$\frac{\partial U}{\partial t} = D\nabla^2 U + \gamma \left[ \frac{pnU^2}{V(1+mU^2)} - cU \right] \quad (1)$$

$$\frac{\partial V}{\partial t} = \nabla^2 V + \gamma \left[ bnU^2 - eV \right] \quad (2)$$

$$\frac{\partial n}{\partial t} = q\nabla^2 n - \chi \left[ \nabla \cdot \left( \frac{n}{(k_n + U)^2} \nabla U \right) \right] + r_n n(1-n) \quad (3)$$

In Eq.(1)-(3), U, V, and n are dimensionless concentrations of activator (U), inhibitor (V), and density of the water molecules (n) as functions of space coordinate (x, y, z) and time (t); c and e are the degradation rate of activator (C-AM) and inhibitor (RT-AM). The temperature difference between an arbitrary region and the surrounding regions drives the thermal energy quantum to diffuse. The diffusion and propagation of C-AMs, RT-AMs and H$_2$O are modelled by the terms of $D\nabla^2 U$, $\nabla^2 V$ and $q\nabla^2 n$, which mathematically serve as the dynamic forms describing the physical transfer of thermal energy in this system. D and q are the ratios of diffusion coefficients for activator-to-inhibitor, and water molecules-to-inhibitor, respectively.

During the cooling process of the freezer, more and more cold air was produced when heat was physically absorbed on the evaporator of the refrigerator. In the according mathematical model, the C-AM density becomes higher when more cold air is produced. The regions filled with C-AMs get cooled, which will get more and more air molecule cooled around. Thus, a term of autocatalytic production of activator (C-AM) is presented in the activator equation. However, the freezer cannot unlimitedly produce the CM because the freezer in reality has an upper limit for its cooling power. Therefore, there is a saturation term of autocatalytic production of activator, e.g. $pnU^2/[V(1+mU^2)]$. the cooled C-AMs enhance the condensation of water vapor and thus the frost and ice appeared. During this condensation process of water vapor, heat is released and thus counteract some C-AMs, which is described by $-cU$. In the eq. (2), the C-AMs can produce RT-AMs because the regions filled with more C-AM will be cooled and thus the frost and ice appeared. During this condensation process of water vapor, heat is released and thus more AMs are generated, and this effect is described by the term $bnU^2$ that contribute the incensement of V over time. In the meantime, the sublimation of the frost and ice also occurred and some heat is physically absorbed during this process, which is described by the term $-eV$. b is the coefficient representing the relative production of inhibitor to activator; $\gamma$ is a scaling factor related to domain size, phase-transition timescale of water, and inhibitor diffusivity;

The condensing and dispersing process of H$_2$O molecules could be modeled by an effect called "thermo-taxis" defined in this study. Here the "cooling-taxis" is defined as the movement of an H$_2$O molecule including condensing or dispersing in response to a stimulus such as the higher or lower thermal energy at local positions. The water molecules at an arbitrary position tend to aggregate and condense at the positions having a negative gradient of CM density. It because the negative gradient of C-AM density indicates a higher density of C-AMs at present position than the adjacent position. The higher density of C-AMs at present position leads to cooling effect and thus leads to the local condensation of the water molecules dispersed in the air. So the density of water molecules at present position becomes higher and the ice crystals would form when the temperature is low enough ( -24 ~ -26℃) in the air. The gradient of

C-AM density ($\nabla U$) drive the water molecules at local positions to aggregate inward or disperse outward, which could be modelled by the divergence of a physical quantity proportional to $\nabla U$, e.g. $\nabla \cdot (K\nabla U)$. Here $K = -\chi n/(k_n + U)^2$ because the density of water molecules ($n$) and the density of C-AMs ($U$) also contribute to thermo-taxis effect of H$_2$O molecules. The larger $n$ will enhance the movement of H$_2$O molecules for aggregating inward or dispersing outward the current point. For instant, the larger $n$ may make the local frost has a larger surface area and enhance the condensation effect. Moreover, the increased $U$ will leads to a saturation of "cooling-taxis" due to the limited capability of U to attract the water vapor to condensation ($\nabla U < 0$) at current position or dispersion outwards ($\nabla U > 0$) from the current position. Therefore, the term can be used to describe the contribution of cooling-taxis effect of water vapor molecules to the term $\partial n / \partial t$.

According to the modelling above, we simulated the final 3D morphology of the self-assembled water molecules and the result is shown in Figure 3, which presents a hollow tube morphology in the air layer of the 3D computation domain that very similar to the emerged ice tubes within the 5-mm-thick layer that sandwiched between the two frosted surfaces in the experiment. The simulated results include two types of the hollow tubes: one is the standalone and approximately parallel straight tubes as shown in Figure 3(a); while the other is the irregular and intersecting tubes. Although the simulated morphologies are not exactly the same with the experimental results, the main features including hollow tubular topology and the intersecting axes of agminated tubes have been modeled and predicted by this mathematic scheme. The parameters for the regions of rubber-frost and foam-frost are assumed $D_F = 0.007$ for C-AMs (activators), and $q_F = 7 \times 10^{-5}$, $\chi_F = 7 \times 10^{-5}$ for H$_2$O molecules in both (a) and (b); the parameters for the air layer are assumed $D_A = 0.011$ in Figure 3a and $D_A = 0.010$ in Figure 3b, and $q_A = 2 \times 10^{-4}$, $\chi_A = 1 \times 10^{-3}$ in Figure 3a and $\chi_A = 2.4 \times 10^{-4}$ in Figure 3b. Other parameters were estimated as $p = 0.7$, $b = 1.0$, $c = 0.034$, $e = 0.02$, $\gamma = 400$, $r_n = 5.6$.

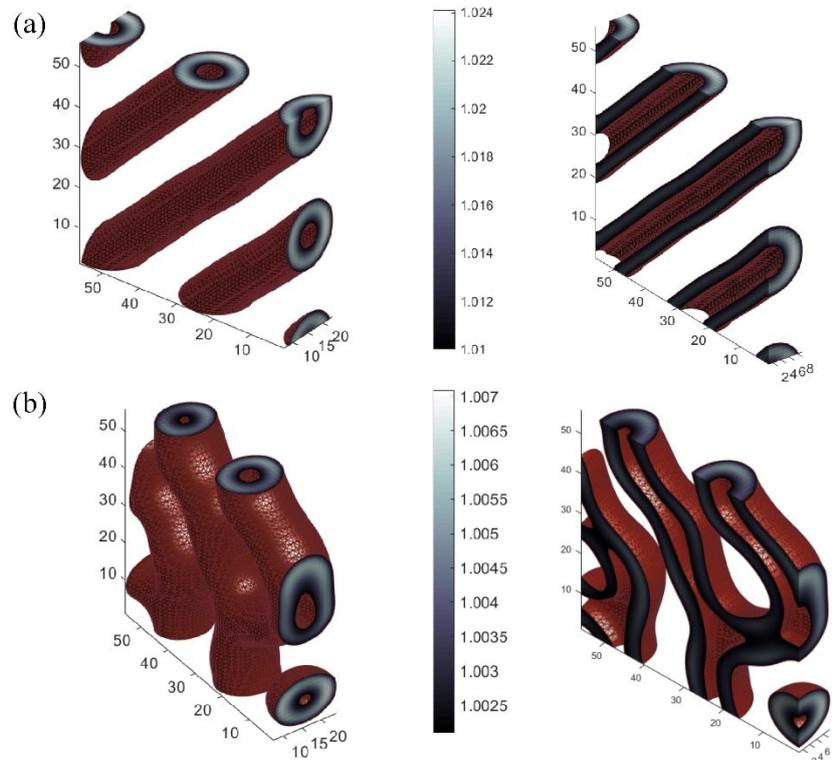

Figure 3. Simulated hollow tubes composed of condensed water molecules within the air layer sandwiched between the two frosted surfaces. The right-half figure is the cutaway view of the corresponding left-half figure to present the lumens inside the simulated tubes. The parameters for the regions of rubber-frost and foam-frost are assumed $D_F = 0.007$ for the C-AMs (activators), and $q_F = 7\times10^{-5}$, $\chi_F = 7\times10^{-5}$ for H$_2$O molecules; the parameters for the air layer is assumed $D_A = 0.011$ in (a) and $D_A = 0.01$ in (b) for the C-AMs (activators), and $q_A = 2\times10^{-4}$, $\chi_A = 1\times10^{-3}$ in (a) and $\chi_A = 2.4\times10^{-4}$ in (b). Other parameters were estimated as $p = 0.7$, $b = 1.0$, $c = 0.034$, $e = 0.02$, $\gamma = 400$, $r_n = 5.6$.

IV. SUMMARY

The hollow ice columns can be self-formed within an 5-mm-thick layer that sandwiched between the two frosted surfaces in a freezer at the temperature of -27 ℃ and the humidity of ~ 95-99%. A simulation model based on Turing instability is built by introducing the cooled air molecule (C-AM) as the activator and room-temperature air molecule (RT-AM), and incorporating the diffusion of C-AMs and RT-AMs, and the heat-transfer mediated condensation of water vapor from the perspective of Turing's reaction-diffusion (RD) process. The simulation results indicate that this modelling framework can generate the hollow tube structures that have similar morphology and topology to the ice tubes formed in the experiment.

ACKNOWLEDGMENT

This work was supported by the National Natural Science Foundation of China (51875170), Fundamental Research Funds for the Central Universities of China (B200202225 and 2018B22414), and Changzhou Sci&Tech Program (CE20195037).